# Local Measurements of Shubnikov-de Haas Oscillations in Graphene Systems

Ya-Ning Ren, Mo-Han Zhang, Chao Yan, Yu Zhang, and Lin He[†]

Center for Advanced Quantum Studies, Department of Physics, Beijing Normal University, Beijing, 100875, People's Republic of China

[†]Correspondence and requests for materials should be addressed to L.H. (e-mail: helin@bnu.edu.cn).

**Shubnikov-de Haas (SdH) oscillations, the most well-known magneto-oscillations caused by the quantization of electron energy levels in the presence of magnetic fields in two-dimensional (2D) electron systems, can be used to determine Fermi-surface properties and directly measure the Berry phase of the 2D systems. It is usually thought that transport measurements are required to measure the SdH oscillations. Contradicting this belief, we demonstrate that the SdH oscillations can be measured in graphene systems by carrying out scanning tunneling spectroscopy (STS) measurements. The energy-momentum dispersions and Berry phases of monolayer, Bernal-stacked bilayer, and ABC-stacked trilayer graphene are obtained according to the measured SdH oscillations in the STS spectra. It is possible to obtain the SdH oscillations when the size of the 2D systems is larger than the magnetic length and, importantly, no gate electrode is required in the STS measurement, therefore, the reported method in this work is applicable to a wide range of materials.**

In two-dimensional (2D) electron systems, a perpendicular magnetic field leads to Landau quantization and, consequently, the density of states (DOS) at a fixed energy will show oscillation caused by the quantization of electron energy levels. Figure 1a schematically shows evolution of the DOS of the monolayer graphene as a function of perpendicular magnetic fields. The electronic states condense into the well-defined Landau levels of massless Dirac fermions, which lead to the oscillation of the DOS at a fixed energy. Such oscillatory effect normally is measured through transport measurements. Then, Shubnikov-de Haas (SdH) oscillations in magnetoresistance, an oscillatory dependence of the electrical resistivity on the magnetic field, can be observed [1,2], as schematically shown in Fig. 1b. According to the SdH oscillations, we can extract two fundamental properties in determining the electronic properties of the studied systems, *i.e.*, the energy-momentum dispersion and Berry phase at a single energy at the Fermi surface. Therefore, the SdH oscillations become the most well-known magneto-oscillations to characterize a wide range of 2D systems [3-14]. However, to measure the SdH oscillations, one needs relatively large samples sizes to carry out transport measurements and, additionally, gate electrode is required to change the energy of the sample at the Fermi level, which impose restrictions on the measurements.

On the other hand, scanning tunneling microscopy (STM) measurements could directly probe the DOS of the 2D systems in a wide energy range (Fig. 1c). Therefore, the Landau quantization of the 2D systems induced by the magnetic fields was widely explored by using STM measurements [15-27]. However, the energy-momentum dispersion and, especially, the Berry phase of the studied systems have not been retrieved from such STM measurements. In this Letter, we demonstrate that the SdH oscillations in the 2D systems can be measured in a rather wide energy range by carrying out STM measurements. The SdH oscillations of monolayer, Bernal-stacked bilayer, and ABC-stacked trilayer graphene are directly extracted from scanning tunneling spectroscopy (STS) spectra measured in different magnetic fields. Consequently, the energy-momentum dispersion and the Berry phase of these systems are obtained. Our experiment demonstrates that the reported measurements are free from these limitations of the traditional SdH oscillations in transport measurements and can be extended to study a wide range of 2D systems.

In our experiment, the measurements were carried out in decoupled monolayer, Bernal bilayer, and ABC-stacked trilayer graphene on multilayer graphene (or on graphite), as reported in previous studies (see Supplemental Materials for details) [18-25]. Usually, the large twist angle between the topmost graphene systems and the supporting graphene multilayer (or graphite) substrate enables that the topmost graphene systems are electronically decoupled from the substrate. The decoupled monolayer, Bernal bilayer, and ABC-stacked trilayer graphene on graphene multilayer (or graphite) can be directly identified by their characteristic high-magnetic-field STS, *i.e.*, *dI/dV*, spectra [18-25]. The *dI/dV* spectra recorded in a constant magnetic field reflect the local DOS at a variable tunneling energy $E$, with $E = E_F + eV_B$ (here $E_F$ is the Fermi energy, $V_B$ is the tunneling bias, and $E_F$ corresponds to $V_B = 0$ V). Figure 2a shows representative high-magnetic-field *dI/dV* spectra of a decoupled monolayer graphene, exhibiting well-defined Landau levels of massless Dirac fermions [15-18] (see Supplemental Fig. 1 for the atomic resolution STM image of a decoupled monolayer graphene). Peaks in the *dI/dV* spectrum occur at magnetic fields where energies of the Landau levels $E_n = eV_B$. In pristine monolayer graphene, the energies of the Landau levels $E_n$ depend on the square root of both level index $n$ and magnetic field $B$, $E_n = \text{sgn}(n)V_F\sqrt{2e\hbar|n|B} + E_0$, $n = ...-2, -1, 0, 1, 2...$ . The fit of the experimental data to the Landau quantization of massless Dirac fermions in monolayer graphene yields a Fermi velocity of $V_F = 1.16 \times 10^6$ m/s and $E_0 = E_D = 48$ meV ($E_D$ is the energy of the Dirac point in the monolayer graphene), as shown in Fig. 2b.

Below we will demonstrate that the STM measurement not only can measure the local DOS as a function of energy (at a fixed $B$), but also can detect the local DOS as a function of magnetic field (at a fixed energy). By measuring conventional *dI/dV* spectra at different magnetic fields with a small interval of the magnetic field $\Delta B$, as shown in Fig. 2a, we can obtain the local DOS in the 2D magnetic and energy, ($B$, $E$), plane. In our experiment, the measurement was performed with the interval of the magnetic field

Δ$B$ = 0.05 T. Then, we obtained the energy-fixed DOS as a function of magnetic field, as shown in Fig. 2c, according to the local DOS in the ($B$, $E$) plane. The largest oscillations originate from the Landau levels sweeping through the fixed energy. Obviously, the observed oscillations of the DOS are quite similar as the SdH oscillations obtained in conventional transport measurements. The main difference is that we can measure the oscillations of the DOS at a variable tunneling energy in the STM measurements rather than at a single energy, *i.e.*, at the Fermi energy, in the transport measurements. Therefore, no gate electrode is required to obtain the band structure properties of both the occupied and empty electronic states in the monolayer graphene by using the STM measurements.

To further explore the oscillations of the DOS in the STM measurements, we plot the SdH-like fan diagram, in which the values of 1/$B_n$ of the *n*th minimum in the DOS are plotted against their index *n*, in Fig. 2d. In graphene systems, the wave function of the quasiparticles can be described by a pseudo-spinor and its chirality determines the zero inverse-field intercept of each line in the fan plot. The resulting fan plot intercepts of the graphene monolayer in our experiment all are very close to 0.5 (or -0.5), as shown in inset of Fig. 2d, which is the same as that observed in pristine monolayer graphene in previous transport measurements [3,4], indicating a Berry phase of π in the monolayer graphene. Besides the Berry phase, we can also extract the energy-momentum dispersion from the SdH-like fan diagram. The frequency of oscillation in 1/$B$ at energy $E$ can be given by $B_E = (\hbar/2\pi e)A_E$, where $A_E$ is the cross-sectional *k*-space area in a plane normal to the magnetic field. With assuming circular constant-energy contours of area $A_E = \pi k_E^2$ in graphene, then the wave vector in monolayer graphene can be written as $k_E = [\left(\frac{4\pi e}{h}\right) B_E]^{1/2}$. Figure 2e shows the linear energy-momentum dispersion determined from the SdH-like fan diagram. A linear fit to the dispersion data yields a Fermi velocity of $V_F$ = 1.15×10$^6$ m/s for both electrons and holes, with the Dirac point $E_D$ = 44 meV, further confirming massless Dirac fermions in the studied graphene system and in good agreement with the values obtained directly from the analysis of the Landau quantization. Obviously, we can obtain the energy-momentum dispersion and Berry phase of the studied systems according to the SdH oscillations in the STM measurements, as only observed in the transport measurements in previous studies.

In our experiment, similar STM measurements were carried out in decoupled Bernal bilayer and ABC-stacked trilayer graphene on graphite (see Supplemental Fig. 1 for the atomic resolution STM image of a decoupled bilayer graphene and trilayer graphene). Figure 3 summarizes a representative result obtained in the Bernal bilayer graphene. The high-magnetic-field spectra exhibit Landau quantization of massive Dirac fermions, as expected for gapped Bernal bilayer graphene (see Supplemental material for details). The fit of the experimental data to the Landau quantization of massive Dirac fermions in bilayer graphene yields the energy gap $E_g$ = 31.2 meV and, simultaneously, the effective masses for electron and hole are obtained as $(0.0352 \pm 0.0004)m_e$ and $(0.0453 \pm 0.0006)m_e$, respectively ($m_e$ is the free-electron mass). These results agree well with that in the Bernal bilayer graphene, as reported in previous studies (see Supplemental Material for details) [18,20,21,24,26]. Similarly, the local DOS of the Bernal bilayer graphene in the 2D ($B$, $E$) plane are measured through the conventional $dI/dV$ spectra at different magnetic fields with a small interval of the magnetic field $\Delta B$ = 0.05 T. Figure 3c shows the representative energy-fixed DOS in the bilayer graphene as a function of magnetic field and Fig. 3d shows the corresponding SdH-like fan diagram. The fan plot intercepts all fall near 1, indicating a Berry phase of $2\pi$ in Bernal-stacked bilayer graphene [5,26,28]. The parabolic energy-momentum dispersion in Bernal-stacked bilayer graphene also can be determined from the SdH-like fan diagram, as shown in Fig. 3e. According to the parabolic energy-momentum dispersion data, the energy gap is obtained as $E_g$ = 33.8 meV and the effective mass of the charge carriers are obtained as $m_e^* = (0.0425 \pm 0.0009)m_e$ for electrons and $m_h^* = (0.0514 \pm 0.0009)m_e$ for holes, which agree with the values obtained from the analysis of the Landau quantization.

Figure 4 summarizes a typical results obtained in the decoupled ABC-stacked trilayer graphene. The quasiparticles in the ABC-stacked trilayer graphene have a chiral degree $l$ = 3 and the low-energy excitations are cubic dispersion [6,24]. The characteristic

Landau quantization of the $l = 3$ quasiparticles is clearly observed in our experiment (Figs. 4a and 4b, see Supplemental Material for details of analysis). The cubic energy-momentum dispersion is also obtained from the SdH-like fan diagram, as shown in Fig. 4e. More importantly, the Berry phase $3\pi$ of the $l = 3$ quasiparticles is directly measured according to the fan plot intercept, 3/2, of the SdH-like oscillations (see Figs. 4c and 4d for details), which is in good agreement with that obtained in previous transport measurements[6]. In our experiment, similar SdH oscillations are obtained in our STM studies in different decoupled monolayer, Bernal bilayer, and ABC-stacked trilayer graphene on multilayer graphene (or on graphite). According to our experiment, such a measurement can be easily extended to the multilayer graphene systems. There are two important advantages in the STM measurement, i.e., its nanometer-scale spatial resolution and no requirement of the gate electrode. The two advantages, as demonstrated here, make this measurement potentially applicable in a wide range of 2D materials.

In summary, the SdH oscillations of monolayer, Bernal-stacked bilayer, and ABC-stacked trilayer graphene are systematically studies via the STM measurements. The energy-momentum dispersion for both the empty and occupied states and the Berry phase of the studied systems are directly obtained. The method of measuring SdH oscillations can be applied to multilayer graphene systems and other 2D materials.

## Acknowledgements

This work was supported by the National Natural Science Foundation of China (Grant Nos. 11974050, 11674029). L.H. also acknowledges support from the National Program for Support of Top-notch Young Professionals, support from "the Fundamental Research Funds for the Central Universities", and support from "Chang Jiang Scholars Program".

# Figures

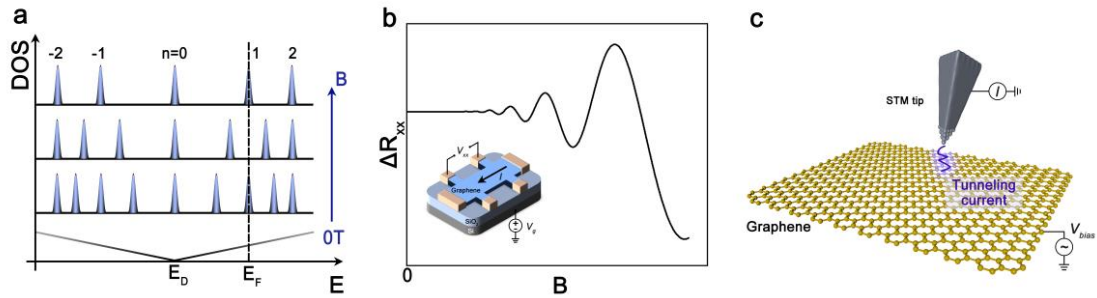

**FIG. 1.** Schematic diagram showing the Shubnikov-de Haas (SdH) oscillations in monolayer graphene. **a.** Schematic images showing the DOS of monolayer graphene in different perpendicular magnetic fields. The peaks correspond to the DOS of the Landau levels. **b.** Schematic SdH oscillations in magnetoresistance measured in transport measurement. Inset: Sketch of a typical experimental device in transport measurements. **c.** Sketch of the STM setup. We can directly probe the DOS of the 2D systems in a wide energy range by using STM.

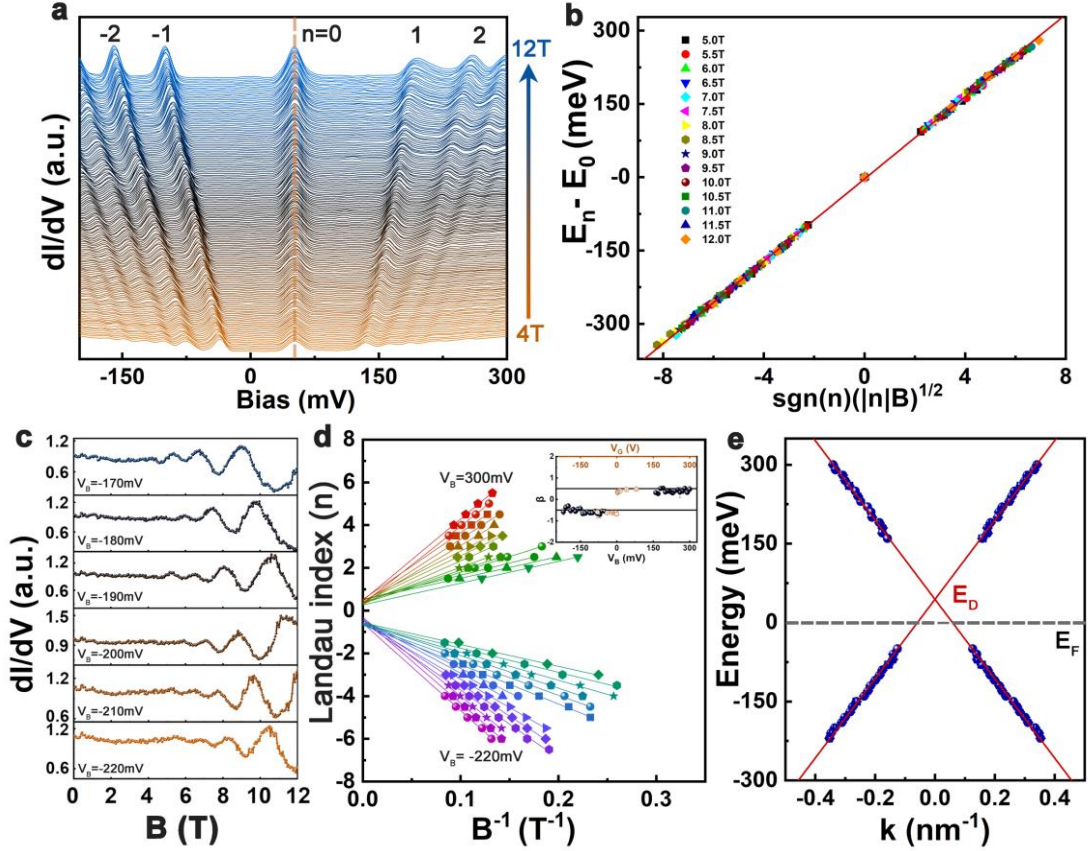

**FIG. 2.** STM measurements of the SdH oscillations in a decoupled monolayer graphene. **a.** Landau level spectra of the monolayer graphene for various magnetic fields from 4 T to 12 T. The data are offset in the *Y* axis for clarity and Landau level indices are labeled. **b.** The Landau level energies taken from panel **a**, show a linear dependence against $sgn(n)(|n|B)^{1/2}$. **c.** The energy-fixed DOS as a function of magnetic field with different tunneling bias $V_B$. **d.** Fan plot showing a linear dependence of the Landau level index from the DOS oscillations in the monolayer graphene on $1/B$ at different tunneling bias. Points of *n* (*n*+1/2) are associated with the *n*th minima (maxima) of the DOS oscillations. The solid lines correspond to the linear fit, in which the slope indicates $B_E$ and the *n*-axis intercept (inset) directly reflects a Berry phase of $\pi$ in the monolayer graphene. **e.** The energy-momentum dispersion of the monolayer graphene obtained from the SdH-like fan diagram. The error is smaller than the symbol size.

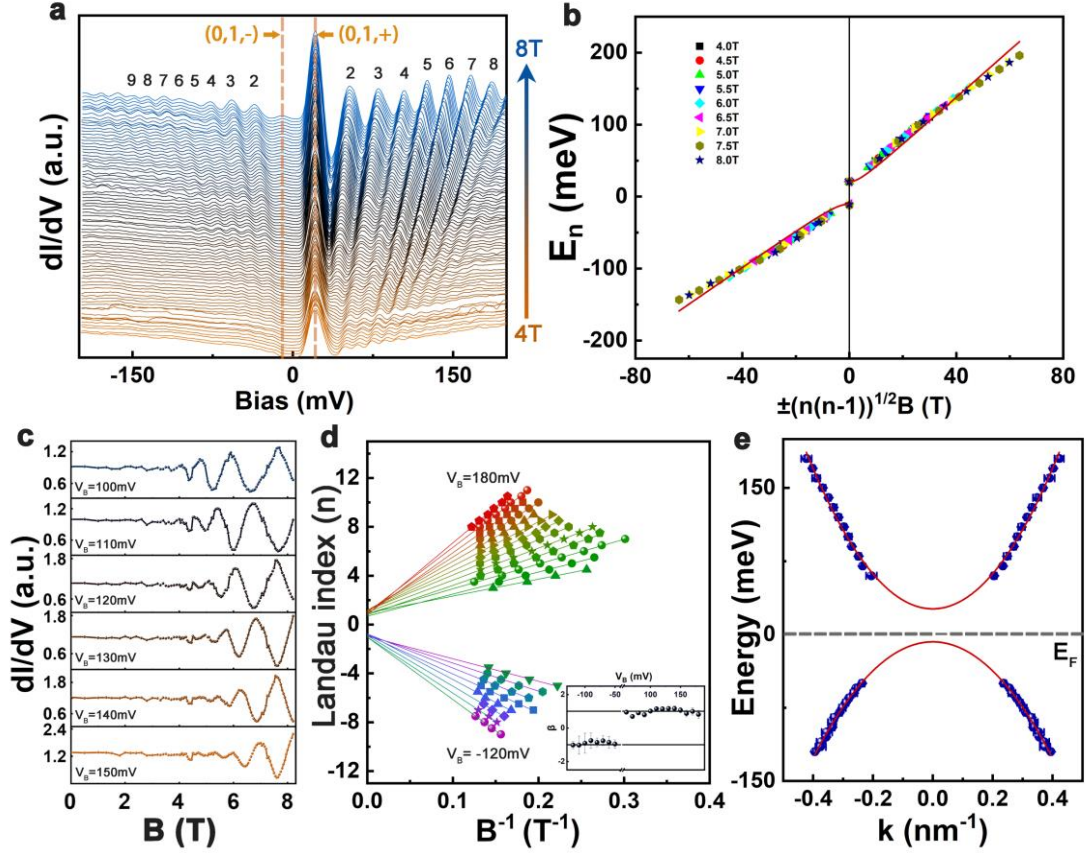

**FIG. 3.** STM measurements of the SdH oscillations in a decoupled Bernal bilayer graphene. **a.** Landau level spectra of the Bernal bilayer graphene for various magnetic fields from 4 T to 8 T. The data are offset in the *Y* axis for clarity and Landau level indices are labeled. **b.** The Landau level energies for different magnetic fields obtained from panel **a** against $\pm(n(n-1))^{1/2}B$. **c.** The energy-fixed DOS as a function of magnetic field with different tunneling bias $V_B$. **d.** A fan diagram of the DOS oscillations in the bilayer graphene at different tunneling bias. Points of *n* (*n*+1/2) are associated with the *n*th minima (maxima) of the oscillations. The solid lines correspond to the linear fit. Inset: The *n*-axis intercept of the linear fit directly reflects a Berry phase of $2\pi$ in the bilayer graphene. **e.** The energy-momentum dispersion of the Bernal bilayer graphene obtained from the SdH-like fan diagram. The error is smaller than the symbol size.

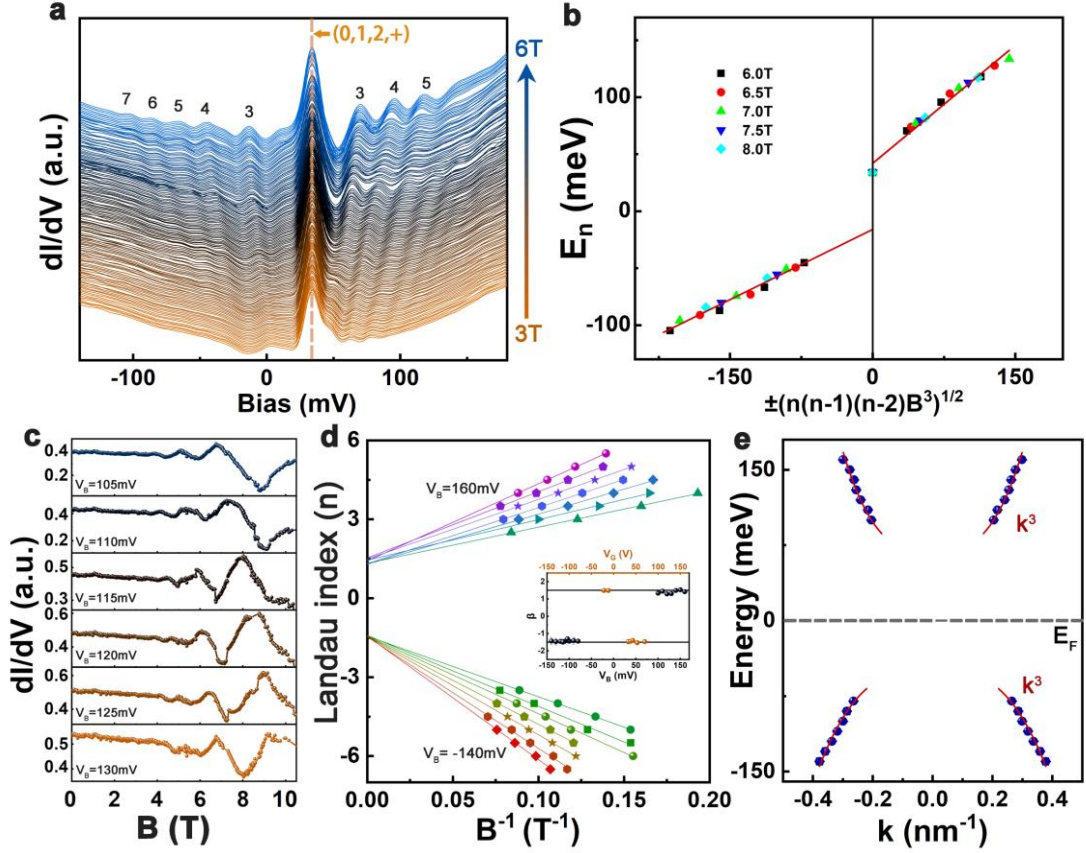

**FIG. 4.** STM measurements of the SdH oscillations in a decoupled ABC-stacked trilayer graphene. **a.** Landau level spectra of the ABC-stacked trilayer graphene for various magnetic fields from 3 T to 6 T. The data are offset in the *Y* axis for clarity and Landau level indices are labeled. **b.** The Landau level energies for different magnetic fields obtained from panel **a** against $\pm(n(n-1)(n-2)B^3)^{1/2}$. **c.** The energy-fixed DOS as a function of magnetic field with different tunneling bias $V_B$. **d.** A fan diagram of the DOS oscillations in the trilayer graphene at different tunneling bias. Points of *n* (*n*+1/2) are associated with the *n*th minima (maxima) of oscillations. The solid lines correspond to the linear fit. Inset: The *n*-axis intercept of the linear fit directly indicates a Berry phase of $3\pi$ in ABC-stacked trilayer graphene. **e.** The energy-momentum dispersion of the ABC-stacked trilayer graphene obtained from the SdH-like fan diagram. The error is smaller than the symbol size.